\documentstyle[pra,aps,psfig,epsf,multicol]{revtex}

\begin{document}
\title{Propagation of entangled light pulses through dispersing
and absorbing channels}
\author{A.V. Chizhov
\footnote{Permanent address:
Joint Institute for Nuclear Research,
Bogoliubov Laboratory of Theoretical Physics,
141980 Dubna, Moscow Region, Russia},
 E. Schmidt, L. Kn\"oll, and D.-G. Welsch}
\address{Friedrich-Schiller-Universit\"{a}t Jena,
Theoretisch-Physikalisches Institut
\\
Max-Wien-Platz 1, D-07743 Jena, Germany}

\address{Received  29 June 2000}
\date{\today}
\maketitle

\begin{abstract}
The problem of decorrelation of entangled
(squeezed-vacuum-type) light pulses of arbitrary shape
passing through dispersive and absorbing four-port devices of
arbitrary frequency response is studied, applying recently
obtained results on quantum
state transformation [Phys. Rev. A {\bf 59}, 4716 (1999)].
The fidelity and indices of (quantum) correlation
based on the von Neumann entropy are calculated,
with special emphasis on the dependence on
the mean photon number, the pulse shape, and the
frequency response of the devices.
In particular, it is shown that
the quantum correlations can decay very rapidly
due to dispersion and absorption, and the degree of
degradation intensifies with increasing
mean photon number.
\end{abstract}

%%%%%%%%%%%%%%%%%%%%%%%%%%%%%%%%%%%%%%%%%%%%%%%%%%%%%%%%%%%%%%%%%%%%%%%%%%%%%

\section{Introduction}
\label{intro}
Quantum-state entanglement of spatially separated systems is one
of the most exciting features of quantum mechanics \cite{Bell87}.
In particular in the rapidly developing field of quan\-tum information
processing (quantum teleportation \cite{Benn93}, quantum cryptography
\cite{Eker91}, and quantum computing \cite{Benn95}), entanglement
has been a subject of intense studies. Recently, continuous-variable
systems have been of increasing interest \cite{Vaid94,Brau98,Furu98,%
Titt98,Titt98a,Titt99,Bren99,Lloyd99,Pari99,vanE99,Loock00,Simon00}.

Since entanglement
is a highly nonclassical property, it is expected to respond very
sensitively to environment influences, and thus it can decrease
very fast. The mechanisms of decoherence are worth to be
studied in detail \cite{Zure91,Zure93}, because they delimit
possible applications.
In particular, optical pulses prepared in entangled states
typically propagate through optical fibers and/or  pass
optical instruments, such as beam splitters, mirrors, and
interferometers. All these devices are built up
by (dielectric) matter that always gives rise to
some dispersion and absorption. As a result, the initially
prepared quantum coherence is destroyed, and the question
arises of what is the characteristic scale of quantum decorrelation.

In this paper the quantum decorrelation of two initially
entangled light pulses that pass through optical devices is
studied. The pulses are regarded as being non\-mo\-nochromatic
modes of arbitrary shape, and the devices are regarded as
being dispersing and absorbing four-port devices of arbitrary
frequency response. The underlying theory of quantum state transformation
has been developed recently \cite{Knoe99}. It is based on a
quantization procedure for the electromagnetic field
in dispersing and absorbing inhomogeneous
dielectrics \cite{Grun95,Matl95,Matl96,Dung98,Sche98},
which is consistent with both the dissipation-fluctuation theorem
and the QED canonical (equal-time) commutation relations.
One interesting aspect of the theory is that given the
complex refractive-index profiles \cite{Grun96} of the devices,
which can be determined  experimentally, the parameters
relevant to the quantum-state transformation can be
calculated without further assumptions.

In Section \ref{se2}
the basic relations for describing the pulse propagation
and the associated quantum-state transformation are given
for the case when the pulses are initially prepared in a
two-mode squeezed vacuum state. The fidelity and  some characteristic
measures of (quantum) correlations of the pulses are
calculated and discussed in Section \ref{se3},
and some concluding remarks are given in Section \ref{se4}.

%%%%%%%%%%%%%%%%%%%%%%%%%%%%%%%%%%%%%%%%%%%%%%%%%%%%%%%%%%%%%%%%%%%%%%%%%%%%%%
%%%%%%%%%%%%%%%%%%%%%%%%%%%%%%%%%%%%%%%%%%%%%%%%%%%%%%%%%%%%%%%%%%%%%%%%%%%%%%

\section{Pulse propagation and quantum-state transformation}
\label{se2}

Let us consider the propagation of two nonclassical light pulses
through lossy four-port devices of given complex refractive-index
profiles \cite{Grun96}. Regarding the pulses as nonmonochromatic
modes, we may define annihilation operators for the two
pulses in terms of the annihilation operators associated with
the monochromatic modes that form the pulses \cite{titu66,Blow90},
\begin{eqnarray}
\label{2.01}
      \hat{a}[\eta] &=&\int_{0}^{\infty } d\omega\,
      \eta ^{*}( \omega) \hat{a}(\omega),
      \\
      \hat{d}[\tilde{\eta}] &=&\int_{0}^{\infty }
      d\omega\, \tilde{\eta} ^{*}(\omega) \hat{d}(\omega),
      \label{2.02}
      \end{eqnarray}
where $\eta(\omega)$ and $\tilde{\eta}(\omega)$
are normalized functions that describe the pulse profiles,
\begin{eqnarray}
\label{2.03}
      \|\eta\|^2\equiv
      ( \eta | \eta ) &\equiv& \int_{0}^{\infty}
      d\omega\, \eta^{*}(\omega) \eta(\omega) = 1 ,
      \\
      \|\tilde{\eta}\|^2\equiv
      ( \tilde{\eta} | \tilde{\eta} ) &\equiv& \int_{0}^{\infty}
      d\omega\, \tilde{\eta}^{*}(\omega)
      \tilde{\eta}(\omega) = 1 .
      \end{eqnarray}
Using the continuous-mode bosonic commutation relations
\begin{equation}
\label{2.04}
      \left[ \hat a(\omega) ,
      \hat a^{\dagger }(\omega')
      \right] =
      \delta(\omega \!-\! \omega') =
      \left[ \hat d(\omega) ,
      \hat d^{\dagger }(\omega')
      \right]
      \end{equation}
(the other commutators being zero), it follows that
\begin{equation}
\label{2.06}
      \left[ \hat a[\eta] , \hat a^{\dagger }[\eta]
      \right] = 1 =
      \left[ \hat d[\tilde{\eta}] , \hat d^{\dagger }[\tilde{\eta}]
      \right].
      \end{equation}

Now let us assume that the pulses are initially prepared
in an entangled quantum state of the type of a two-mode squeezed
vacuum state
\begin{equation}
\label{2.08}|{\Psi}_{\rm in} \rangle
      =\exp\!\left\{ q^{*}\hat{a}[\eta]\,\hat{d}[\tilde{\eta}] -{\rm H.c.}
      \right\} |0\rangle \equiv \hat{S}\left(q\right) |0\rangle ,
      \end{equation}
with $q=|q|e^{i\varphi _q}$ being the squeezing parameter.
Most studies of continuous-variable systems in quantum information
processing have been based on such states 
\cite{Brau98,Furu98,Lloyd99,Loock00}.
After having passed the devices, the output state of the pulses reads as
\begin{equation}
\label{2.09}|{\Psi }_{\rm out}\rangle
=\exp \left\{ q^{*}\hat{a}'[\eta]\,\hat{d}'[\tilde{\eta}] -{\rm H.c.}
      \right\} |0\rangle \equiv \hat{S}'(q) |0\rangle,
      \end{equation}
where
\begin{eqnarray}
\label{2.10}
      \hat{a}'[\eta] &=& \int_{0}^{\infty } d\omega\,
      \eta ^{*}(\omega)
      \hat{a}'(\omega), \\
\label{2.10a}
      \hat{d}'[\tilde{\eta}] &=& \int_{0}^{\infty } d\omega\,
      \tilde{\eta} ^{*}(\omega)
      \hat{d}'(\omega),
      \end{eqnarray}
and the continuous-mode operators $\hat{a}'(\omega )$ and $\hat{d}'(\omega )$
are given by \cite{Knoe99}
\begin{eqnarray}
\label{2.11}
      \hat{a}'(\omega ) &=&
      \sum_{i=1}^{2}\left[
      T_{i1}^{*}(\omega) \hat{a}_i(\omega)
      +F_{i1}^{*}(\omega) \hat g_i(\omega) \right] , \\
\label{2.11a}
      \hat{d}'(\omega ) &=&
      \sum_{i=1}^{2}\left[
      \tilde{T}_{i1}^{*}(\omega) \hat{d}_i(\omega)
      +\tilde{F}_{i1}^{*}(\omega) \hat h_i(\omega) \right]
      \end{eqnarray}
[see \ref{ior}, equation (\ref{A.10})].
In equations (\ref{2.11}), (\ref{2.11a}) the continuous-mode operators
$\hat a_1(\omega ) \equiv \hat a(\omega )$ and
$\hat d_1(\omega ) \equiv \hat d(\omega )$
belong to the fields entering the first input ports of the
two four-port devices, whereas $\hat a_2(\omega )$ and
$\hat d_2(\omega )$ belong to the fields
entering the second input ports,  and the bosonic operators
$\hat{g}_i(\omega )$ and $\hat{h}_i(\omega )$ describe
excitations of the two devices.
The quantities  $T_{11}( \omega ) \equiv T( \omega )$,
$\tilde{T}_{11}( \omega ) \equiv \tilde{T}( \omega )$
and $T_{21}( \omega ) \equiv R( \omega )$,
$\tilde{T}_{21}( \omega ) \equiv \tilde{R}( \omega )$
are respectively the transmission and reflection coefficients
of the four-port devices with respect to the incoming fields at the
first input ports. In what follows we assume that
the second input ports of the devices are unused, that is,
the fields there are in the vacuum state, the corresponding
variables together with the device variables being referred to as the
environment $\cal{E}$. Moreover, we assume that the devices are not
excited.

It is  convenient to represent each of the operators $\hat{a}'[\eta]$
and $\hat{d}'[\tilde{\eta}]$ in equations (\ref{2.10}), (\ref{2.10a})
as a sum of two other independent bosonic operators,
\begin{eqnarray}
\label{2.13}
      \hat{a}'[\eta] &=& \|T\eta\|\,\hat a[\eta']
      + (1-\|T\eta\|^2)^{1/2}\,\hat q_\eta\,, \\
\label{2.13a}
      \hat{d}'[\tilde{\eta}] &=&
      \|\tilde{T}\tilde{\eta}\|\,\hat d[\tilde{\eta}']
      + (1-\|\tilde{T}\tilde{\eta}\|^2)^{1/2}\,\hat p_{\tilde{\eta}}\,,
      \end{eqnarray}
where
$\eta'(\omega )= T(\omega)\eta(\omega)/\|T\eta\|$
and $\tilde{\eta}'(\omega )= \tilde{T}(\omega )\tilde{\eta}(\omega )/
\|\tilde{T}\tilde{\eta}\|$. Note that the
operators $\hat q_\eta$ and $\hat p_{\tilde{\eta}}$ belong to
the environment $\cal{E}$,
\begin{eqnarray}
\label{2.14}
      (1-\|T\eta\|^2)^{1/2}\,\hat q_\eta &=&
       \hat{a}_2[R\eta]
      +\sum_{i=1}^{2}
      \hat g_i[F_{i1}\eta], \\
      (1-\|\tilde{T}\tilde{\eta}\|^2)^{1/2}\,\hat p_{\tilde{\eta}} &=&
       \hat{d}_2\!\left[\tilde{R}\tilde{\eta}\right]
      +\sum_{i=1}^{2}
      \hat h_i\!\left[\tilde{F}_{i1}\tilde{\eta}\right].
      \end{eqnarray}

Given the output quantum state $|\Psi_{\rm out}\rangle$ of the system, the
(symmetric) characteristic function
\begin{equation}
\label{2.17}
      \Phi_{\rm out} (\alpha ,\beta )=\left\langle \exp\! \left( \alpha
      \hat a^{\dagger }[\eta'] +
      \beta \hat d^{\dagger }[\tilde{\eta}']
      -{\rm H.c.}\right) \right\rangle _{\rm out}
      \end{equation}
can be calculated.
In \ref{app.chi2} it is shown that
\begin{eqnarray}
\label{2.19}
\lefteqn{
      \Phi_{\rm out} \left( \alpha ,\beta \right) =
      \exp \Big\{
      -{\textstyle {1 \over 2}}\left(\left| \alpha \right| ^2
      \left[ 1+\left( \cosh 2|q|-1\right)
      \|T\eta\|^2\right] \right.
}
\nonumber\\&&\hspace{2ex}
      +\left.
      \left| \beta \right| ^2\left[ 1+\left( \cosh 2|q|-1\right)
      \|\tilde{T}\tilde{\eta}\|^2\right] \right)
\nonumber\\&&\hspace{2ex}
      -{\textstyle {1 \over 2}}\left(\alpha
      \beta e^{-i\varphi _q}\sinh 2|q|+
      \alpha ^{*}\beta ^{*}e^{i\varphi _q}\sinh 2|q|\right)\|T\eta\|
      \|\tilde{T}\tilde{\eta}\|
      \Big\}.
      \end{eqnarray}
Note that
$\Phi_{\rm out}( \alpha ,\beta )$ is a function only of the moduli
$|T(\omega )|$ and $|\tilde{T}(\omega )|$ of the
trans\-mission coefficients.
For $T(\omega) = \tilde{T}(\omega) = 1$,
the characteristic function of the incoming fields is recognized.

%%%%%%%%%%%%%%%%%%%%%%%%%%%%%%%%%%%%%%%%%%%%%%%%%%%%%%%%%%%%%%%%%%%%%%%%%%%%%%
%%%%%%%%%%%%%%%%%%%%%%%%%%%%%%%%%%%%%%%%%%%%%%%%%%%%%%%%%%%%%%%%%%%%%%%%%%%%%%

\section{Pulse correlations}
\label{se3}

A measure of the entanglement of two subsystems
$(A)$ and $(B)$ of a composed system $(AB)$ that is prepared
in some mixed state $\hat{\varrho}$
is the quantum relative entropy $E$ \cite{Vedr98}, the quantum analog
of the classical Kullback-Leibler entropy,
\begin{eqnarray}
\label{2.19a}
     E(\hat{\varrho })=\min_{\hat{\sigma} \mit\epsilon {\cal S}}\,
     {\rm Tr}\,[\hat{\varrho }\,(\ln{\hat{\varrho }}\!
     -\!  \ln{\hat{\sigma}})],
      \end{eqnarray}
where ${\cal S}$ is the set of all
separable quantum states the composed system can be prepared in.
For pure states, $\hat{\varrho }$ $=$ $|\Psi\rangle\langle\Psi|$,
the entanglement measure (\ref{2.19a}) reduces to the
von Neumann entropy of one subsystem
\begin{eqnarray}
\label{2.19b}
     E(\hat{\varrho}) = S_A
     = {\rm Tr}^{(A)}\Big(\hat{\varrho } ^{(A)} \ln{\hat{\varrho }
     ^{(A)}}\Big) =S_B,
     \qquad
     \hat{\varrho } ^{(A)} = {\rm Tr}^{(B)}\hat{\varrho }
      \end{eqnarray}
[Tr$^{(A)}$  (Tr$^{(B)}$), trace with respect to the subsystem A (B)].
In the case where $|\Psi\rangle$ is the
two-mode squeezed vacuum state (\ref{2.08}) it follows that
\begin{eqnarray}
\label{2.19c}
      E(\hat{\varrho}) = S_{\rm th}(\bar{n}_{\rm sq}) =
      (\bar{n}_{\rm sq}+1)\,\ln{(\bar{n}_{\rm sq}+1)}-
      \bar{n}_{\rm sq}\,\ln{\bar{n}_{\rm sq}}\,,
      \end{eqnarray}
with $\bar{n}_{\rm sq}={\sinh }^2|q|$ being the mean number of photons
in each mode. Hence, the entang\-lement is given by the von Neumann entropy
of a thermal state of mean photon number $\bar{n} = \bar{n}_{\rm sq}$.
Note that for large mean photon numbers the entanglement
increases linearly with the squeezing parameter,
$E \approx \ln{\bar{n}_{\rm sq}} \approx 4|q| $.

Unfortunately, there has been no explicit expression for
calculating the entanglement measure (\ref{2.19a}) of mixed states,
which are typically observed in noisy systems.
Extensive numerical procedures would be in\-dispen\-sable in general,
which dramatically grow up with increasing dimension of the
Hilbert space \cite{SKOW00}. Therefore some other correlation
measures have been introduced. Although they are not purely quantum
correlation measures [as is the entanglement measure (\ref{2.19a})],
they may be very helpful to gain insight into the problem
of degradation of quantum correlations.

%%%%%%%%%%%%%%%%%%%%%%%%%%%%%%%%%%%%%%%%%%%%%%%%%%%%%%%%%%%%%%%%%%%%%%%%%

\subsection{Fidelity}

\label{fidelity}

Fidelity can be regarded as being a measure of how close to each
other are two (system) states in the corresponding Hilbert space.
Fidelities have been used, e.g., to describe decoherence effects in
the transmission of quantum information through noisy channels
\cite{Schu96,Barn98} and to characterize the quality of
quantum teleportation \cite{Brau98}.

When $\hat{\varrho}_{\rm in} = |{\Psi}_{\rm in} \rangle \langle
{\Psi}_{\rm in}|$ and $\hat{\varrho}_{\rm out} =
|{\Psi}_{\rm out} \rangle \langle {\Psi}_{\rm out}| $ are respectively
the input and the output density operators of the overall system (composed
of the two pulses and the environment), then the density operators
of the incoming and outgoing fields are respectively
\begin{equation}
\label{3.01}
      \hat{\varrho}_{\rm in}^{(a,d)}
      = {\rm Tr}^{(\cal{E})}\hat{\varrho}_{\rm in}
      \end{equation}
and
\begin{equation}
\label{3.02}
      \hat{\varrho}_{\rm out}^{(a,d)}
      = {\rm Tr}^{(\cal{E})}\hat{\varrho}_{\rm out}\,,
      \end{equation}
(Tr$^{(\cal{E})}$, trace with respect to the environment).
Following c, the fidelity
\begin{equation}
\label{3.03}
      F_e={\rm Tr}\!\left(\,\hat{\varrho}_{\rm in}^{(a,d)}
      \hat{\varrho}_{\rm out}^{(a,d)}\right)
      \end{equation}
can be defined.
In \ref{app.F} it is shown that applying the
quantum state trans\-formation outlined in Section \ref{se2} leads to
\begin{eqnarray}
\label{3.04}
      F_e =
      \left| 1+\left[1-(\eta | T \eta ) (\tilde{\eta} |
      \tilde{T} \tilde{\eta} ) \right]
      \,\bar{n}_{\rm sq} \right| ^{-2}.
      \end{eqnarray}
Note that $F_e$ in equation (\ref{3.04}) refers to the transmitted
light. Replacing $T(\omega)$ and $\tilde{T}(\omega)$ with
$R(\omega)$ and $\tilde{R}(\omega)$ respectively, the fidelity with
respect to the reflected fields is obtained.
From equation (\ref{3.04}) it is seen that the fidelity sensitively
depends on the spectral overlaps of the transmitted and the incoming
pulses, and these overlaps are substantially determined by the
dependence on frequency of the transmission coefficients $T(\omega)$
and $\tilde{T}(\omega)$ (cf. Fig.~\ref{fig4}).
It is worth noting that the fidelity decreases rapidly with increasing
initial mean photon number, that is, with increasing initial
squeezing and thus increasing initial entanglement.

\vspace*{-1.4cm}
\begin{figure}[h]  %%% old %%% 170 530 475 775
\centerline{\psfig{file=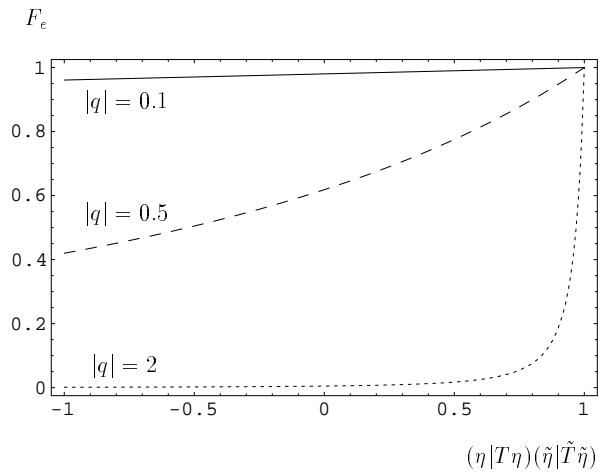,width=8cm,bbllx=170pt,bblly=460pt,%
bburx=475pt,bbury=775pt}}
\caption{\label{fig1}
The fidelity \protect$F_e$, equation (\protect\ref{3.04}),
is shown as a function of (real) $(\eta |T \eta )(\tilde{\eta}
|\tilde{T} \tilde{\eta} )$
for various values of the squeezing parameter $|q|$.
}
\end{figure}

To illustrate the effect, $F_e$ is shown in Fig.~\ref{fig1}
as a function of the product of the overlaps
$(\eta |T \eta ) (\tilde{\eta} |\tilde{T} \tilde{\eta} )$
for different values of the squeezing parameter $|q|$. In the figure
$(\eta |T \eta ) (\tilde{\eta} |\tilde{T} \tilde{\eta} )$ is assumed
to be real. If the phases of $T(\omega)$ and $\tilde{T}(\omega)$
did not depend on $\omega$, then real $(\eta |T \eta )
(\tilde{\eta} |\tilde{T} \tilde{\eta} )$ would
correspond to the maximally attainable fidelity
(which could be also called entanglement fidelity~\cite{Plen98}).

In quantum teleportation a strongly entangled two-mode
squeezed vacuum is desired. Since in this case the photon number
must be large, the state should tend to a macroscopic (at least
mesoscopic) state, and hence its nonclassical features can become
extremely unstable. Clearly, in the case of discrete-variable
systems the entanglement must not necessarily increase with the
mean photon number.
For example, in the case of a two-mode state of the type
\begin{equation}
\label{3.04aa}
      |{\Psi}_{\rm in} \rangle =
      \frac{1}{\sqrt{1\!+\!|\lambda|^2}}
      \left(|00\rangle \!+\! \lambda |nn\rangle\right) =
      \frac{1}{\sqrt{1\!+\!|\lambda|^2}}
      \left[1 \!+\! \lambda \,
      \frac{\left(\hat{a}[\eta]\,\hat{d}[\tilde{\eta}]\right)^n}{n!}
      \right] |0\rangle\,,
      \end{equation}
the entanglement is
\begin{equation}
\label{3.04ab}
E = \frac{S_{\rm th}(|\lambda|^2)}{1+|\lambda|^2},
\end{equation}
whereas the mean photon number in one mode
is $\bar{n}= n |\lambda|^2/(1+|\lambda|^2)$.
The entanglement attains its maximal value $E = \ln 2$
at $|\lambda| = 1$, which corresponds to a Bell-type state.
Nevertheless, when $\bar{n}$ becomes large the entanglement
of the transmitted field decreases exponentially with $\bar{n}$
\cite{SKOW00}. A similar behaviour is observed for the fidelity.
Using the results in Ref.~\cite{SKOW00}, the fidelity (with respect
to the transmitted fields) is obtained to be
\begin{eqnarray}
\label{3.04ad}
\lefteqn{
      F_e =
      \frac{1}{(1+|\lambda|^2)^2}
      \left[\left| 1+|\lambda|^2(\eta | T \eta )^n (\tilde{\eta}
      | \tilde{T} \tilde{\eta} )^n
      \right|^2 \right.
}
\nonumber\\&&\hspace{6ex}
      + \left.
      |\lambda|^2\left(1-\|T\eta\|^{2}\right)^n
      \left(1-\|\tilde{T}\tilde{\eta}\|^{2}\right)^n\right].
      \end{eqnarray}
It is seen that with increasing value of $n$ the fidelity rapidly
decreases to the minimal value, % of $F_e = (1+|\lambda|^2)^{-2}$,
i.e., $F_e = 0.5$ for $|\lambda| = 1$.

%%%%%%%%%%%%%%%%%%%%%%%%%%%%%%%%%%%%%%%%%%%%%%%%%%%%%%%%%%%%%%%%%%%%%%%%%%%%%%%

\subsection{Entropic correlation measures}
\label{entanglement}

Correlation measures can be defined employing the von
Neumann entropy. A very general correlation measure
is the index of correlation \cite{Barn89}
\begin{equation}
      \label{3.06}
      I_c=S_a+S_d-S_{ad}\,,
\end{equation}
where
\begin{equation}
\label{3.08}
      S_{ad}=-{\rm Tr}\left(\hat{\varrho}_{\rm out}^{(a,d)}
      \ln \hat{\varrho}_{\rm out}^{(a,d)} \right)
\end{equation}
is the entropy of the two-pulse system,
and
\begin{equation}
\label{3.07}
      S_m=-{\rm Tr}\left( \hat{\rho }_{\rm out}^{(m)}
      \ln \hat{\rho }_{\rm out}^{(m)}\right)
\end{equation}
($m=a,d$) are the entropies of the single-pulse systems,
\begin{equation}
\label{3.04a}
      \hat{\varrho }_{\rm out}^{(a)}
      = {\rm Tr}^{(d)}\,\hat{\varrho}_{\rm out}^{(a,d)},
      \qquad
      \hat{\varrho }_{\rm out}^{(d)}
      = {\rm Tr}^{(a)}\hat{\varrho}_{\rm out}^{(a,d)}.
      \end{equation}
Note that $I_c$ is bounded from below by $|S_a-S_d|$
and from above by \mbox{$S_a$ $\!+$ $\!S_d$}. Further, $I_c$
is an upper bound of the entanglement, $E \leq I_c$,
because $\hat{\varrho }_{\rm out}^{(a)}\otimes
\hat{\varrho }_{\rm out}^{(d)}$ is an element of the set of
separable states ${\cal S}$ of the two-pulse system.

Correlation measures that can be used to formulate
criteria of nonclassical correlation are \cite{Schu96,Barn98}
\begin{eqnarray}
\label{3.09}
      I_e^{(m)} = S_m - S_{ad}
      \end{eqnarray}
($m=a,d$).
Since the entropy of a classical system
must not be less than the entropy of one of its subsystems,
positive values of $I_e^{(m)}$ indicate nonclassical correlation.
Thus, positive values of $I_e^{(a)}$ and/or $I_e^{(b)}$ may be
regarded as indicating entanglement.

As can be seen from equation (\ref{2.19}),
the characteristic function of $\hat{\varrho}_{\rm out}^{(a,d)}$
is of Gaussian type. The same is true for the characteristic function of
$\hat{\varrho }_{\rm out}^{(a)}$ and $\hat{\varrho }_{\rm out}^{(d)}$,
which follows from equation (\ref{2.19}) for $\beta=0$
and $\alpha=0$ respectively. Hence, the entropies $S_m$,
equation (\ref{3.07}), and [after diagonalizing the quadratic form
in the exponent in equation (\ref{2.19})] the entropy $S_{ad}$,
equation (\ref{3.08}), can be obtained analytically in the form
of the entropy of thermal states:
\begin{equation}
\label{3.14}
      S_{a}= S_{\rm th}(n_a),
      \quad
      n_{a}= \|T\eta\|^2 \bar{n}_{\rm sq}\,,
\end{equation}
\begin{equation}
\label{3.15}
      S_{d}= S_{\rm th}(n_d),
      \quad
      n_{d}= \|\tilde{T}\tilde{\eta}\|^2 \bar{n}_{\rm sq}\,,
\end{equation}
\begin{equation}
\label{3.16}
      S_{ad}= S_{\rm th}(n_{ad}),
      \quad
      n_{ad}= (1-\|T\eta\|^2 \|\tilde{T}\tilde{\eta}\|^2) \bar{n}_{\rm sq}\,.
\end{equation}
Equations (\ref{3.14}) -- (\ref{3.16}) again reveal
the typical dependence on the spectral overlaps of the
transmitted and the incoming pulses, the overlaps
being substantially determined by the frequency response of
the devices. Note that only the
absolute values of $T(\omega)$ and $\tilde{T}(\omega)$ appear.
Obviously, the entropies of the reflected light are obtained
by replacing $T(\omega)$ and $\tilde{T}(\omega)$
with $R\left(\omega \right)$ and $\tilde{R}\left(\omega \right)$
respectively.

\vspace*{-0.2cm}
\begin{multicols}{2}

\begin{minipage}{3.2in}
\begin{figure}[h] %%% old %%% 154 534 450 745
\psfig{file=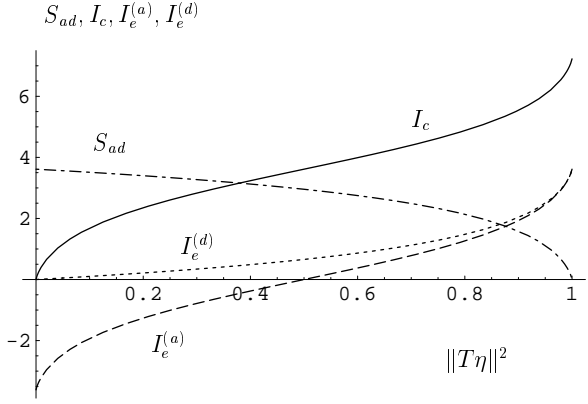,width=8cm,bbllx=154pt,bblly=523pt,%
bburx=450pt,bbury=745pt}
\vspace*{-0.2cm}
\caption{\label{fig2}
The entropy $S_{ad}$, equation (\protect\ref{3.08}),
and the correlation indices $I_c$, equation (\protect\ref{3.06}),
and $I_e^{(a)}$ and $I_e^{(d)}$, equation (\protect\ref{3.09}),
are shown as functions of $\|T\eta\|^2 = (T \eta | T \eta )$
for $|q|=2$ [$\tilde{T}(\omega )=1$].
}
\end{figure}
\end{minipage}

\begin{minipage}{3.2in}
\begin{figure}[h] %%% old %%% 154 534 450 745
\psfig{file=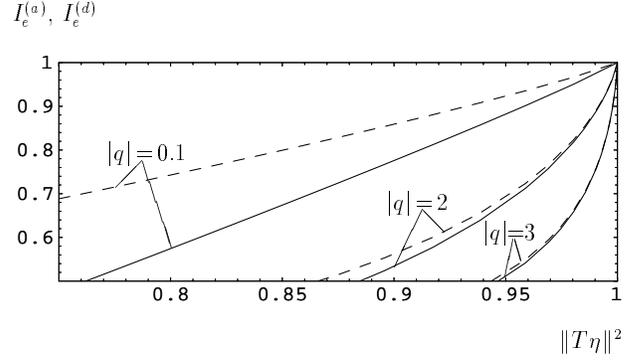,width=8cm,bbllx=185pt,bblly=500pt,bburx=480pt,%
bbury=710pt}
\caption{\label{fig3}
The normalized correlation indices $I_e^{(a)}$ (solid line)
and  $I_e^{(d)}$ (dashed line),
equation (\protect\ref{3.09}), are shown as functions of
$\|T\eta\|^2 = (T \eta | T \eta )$ for various values of
the squeezing parameter $|q|$ [$\tilde{T}(\omega )=1$]. The
values $|q|=0.1$, $|q|=2$, and $|q|=3$, respectively, correspond
to the mean photon numbers $\bar{n}_{\rm sq}= 0.01$,
$\bar{n}_{\rm sq}= 13$, and $\bar{n}_{\rm sq}= 100$.
}
\end{figure}
\end{minipage}

\end{multicols}

Examples of the entropy $S_{ad}$ and the correlation
indices $I_c$ and $I_e^{(m)}$ as functions of $\|T\eta\|^2$
are shown in Fig.~\ref{fig2} for $\|\tilde{T}\tilde{\eta}\|^2 = 1$
(that is, only one channel is noisy).
It is seen that for not too small values of $\|T\eta\|^2$
both $I_e^{(a)}$ and $I_e^{(d)}$ are positive and thus
indicate entanglement. With decreasing value of $\|T\eta\|^2$
they decrease in a similar way as $I_c$.
Whereas $I_e^{(a)}$ referring to the noise channel
becomes negative and thus attains classically allowed values,
$I_e^{(d)}$ referring to the unperturbed channel
remains always positive but becomes small.
The results are in agreement
with the Peres-Horodecki separability criterion for
bipartite Gaussian quantum states \cite{Simon00}. It tells us
that in the low-temperature limit the two-pulse system under
consideration remains inseparable for all values of
$\|T\eta\|^2$ and $\|\tilde{T}\tilde{\eta}\|^2$
(cf. \cite{SOW}).
In order to get insight into the influence of the initial
mean photon number $\bar{n}_{\rm sq}={\sinh }^2|q|$
on the degradation of the quantum correlation,
we have plotted in Fig.~\ref{fig3} the dependence on
$\|T\eta\|^2$ of the correlation indices $I_e^{(a)}$ and $I_e^{(d)}$
for different values of $|q|$.
It is clearly seen that the larger $\bar{n}_{\rm sq}$ becomes,
the faster $I_e^{(a)}$ and $I_e^{(d)}$ decrease.

%%%%%%%%%%%%%%%%%%%%%%%%%%%%%%%%%%%%%%%%%%%%%%%%%%%%%%%%%%%%%%%%%%%%%%%%%%

\subsection{Example: absorbing dielectric plate}
\label{plate}

The values of $(\eta |T \eta )$, $(\tilde{\eta} |\tilde{T}
\tilde{\eta} )$, and $\|T\eta\|^2$, $\|\tilde{T}\tilde{\eta}\|^2$
depend on the chosen pulse forms and on the dependence on
frequency of the transmission coefficients of the devices
$T(\omega)$ and $\tilde{T}(\omega)$.
When the bandwidths of the pulses are sufficiently small,
so that the variation of $T(\omega)$ and $\tilde{T}(\omega)$
within the pulse bandwidths can be disregarded, then
$T(\omega)$ and $\tilde{T}(\omega)$ can be taken
at the mid-frequencies $\omega_f$ and $\tilde{\omega}_f$ of
the pulse, i.e.,
$(\eta |T_{ij} \eta ) \approx T_{ij}(\omega_f)$
and $(\tilde{\eta} |\tilde{T}_{ij} \tilde{\eta} )
\approx \tilde{T}_{ij}(\tilde{\omega}_f)$. In this case,
the results become independent of the pulse shape and
solely reflect the effect of the devices.

\begin{figure}[h]  %%old%% 40 185 375 505
\psfig{file=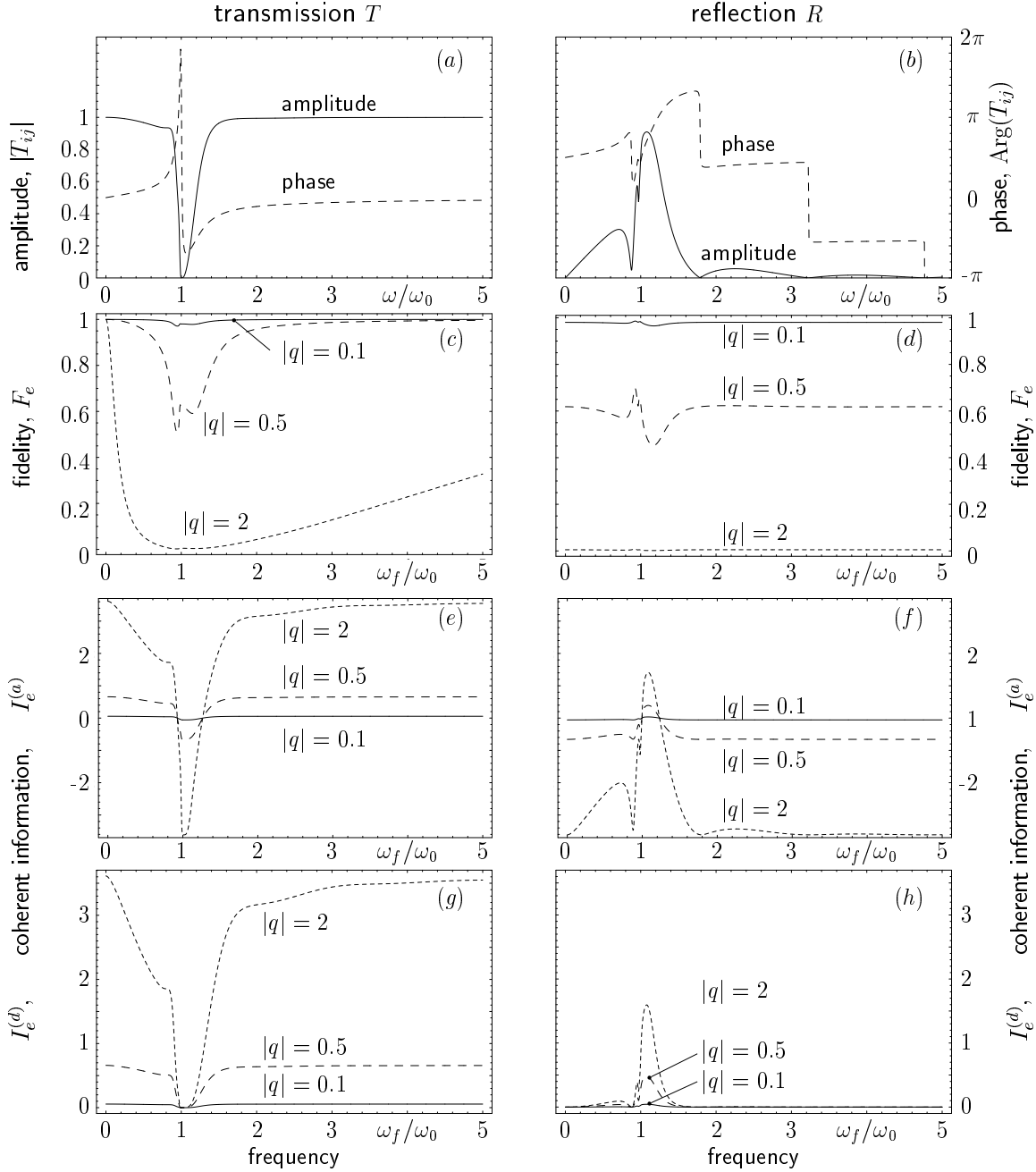,width=6in,bbllx=50pt,bblly=56pt,%
bburx=565pt,bbury=665pt}
\caption{\label{fig4}
The fidelity, equation (\protect\ref{3.04}),
and the correlation indices  $I_e^{(a)}$ and
$I_e^{(d)}$, equation (\protect\ref{3.09})
are shown as functions of the mid-frequency $\omega_f$
of narrow-bandwidth pulses that are transmitted (left figures)
and reflected (right figures) at a dielectric plate
for various values of the squeezing parameter $|q|$
[$\tilde{T}(\omega )=1$].
For comparison, the dependence on frequency of the amplitude
and the phase of the transmission coefficient $T(\omega)$
and the reflection coefficient $R(\omega)$ of the
dielectric plate [complex permittivity
(\protect\ref{2.80}) with $\epsilon_{\rm s}$ $\!=$ $\!1.5$,
$\gamma/\omega_0$ $\!=$ $\!0.01$;
thickness $2c/\omega_0$] are shown.
}
\end{figure}

Let us again consider the case where one pulse passes
through free space and assume that the other pulse passes
through a dielectric plate
whose complex (Lorentz-type) permittivity is given by
\begin{equation}
\label{2.80}
\epsilon = 1+\frac{\epsilon_{\rm s}-1}{1-(\omega/\omega_0)^2-2i
\gamma \omega/\omega_0^2} \,.
\end{equation}
Explicit expressions for the transmission coefficient $T(\omega)$
and the reflection coefficient $R(\omega)$ of such a device
are given in Ref.~\cite{Grun96}.
In Figs.~\ref{fig4}($a$) and ($b$) they are
shown as functions of $\omega$ for
$\epsilon_{\rm s}$ $\!=$ $\!1.5$, $\gamma/\omega_0$ $\!=$ $\!0.01$,
and plate thickness $2c/\omega_0$.
Typically, the transmission amplitude shows a dip near the medium
resonance $\omega_0$, whereas the reflection is peaked there.
In Figs.~\ref{fig4}($c$) and ($d$),
the fidelities $F_e$ of the transmitted and reflected fields
are shown as functions of $\omega_f$. Finally,
Figs.~\ref{fig4}($e$) -- ($h$) show the dependence on $\omega_f$
of the corresponding correlation indices $I_e^{(m)}$.
In particular, it is seen that for transmission $F_e$ and $I_e^{(m)}$
are strongly reduced near the medium resonance,
$\omega_f \approx \omega_0$,
whereas for reflection they are enhanced in that region.

%%%%%%%%%%%%%%%%%%%%%%%%%%%%%%%%%%%%%%%%%%%%%%%%%%%%%%%%%%%%%%%%%%%%
%%%%%%%%%%%%%%%%%%%%%%%%%%%%%%%%%%%%%%%%%%%%%%%%%%%%%%%%%%%%%%%%%%%%

\section{Conclusions}
\label{se4}

We have studied the problem of quantum correlations in a system
of two entangled optical pulses of arbitrary shape,
propagating through dispersing and obsorbing
four-port devices, which may serve as models
of beam splitters, fibres, interferometers etc.
Given the complex refractive-index profiles of the devices,
which can be determined  experimentally, the relevant
parameters for the pulse propagation and the
associated quantum-state transformation can be
calculated without further assumptions. The
devices can be viewed as realizations of noisy
channels for quantum information transmission.

In the calculations we have assumed that the
pulses are initially prepared in a two-mode squeezed
vacuum, which is typically considered in quantum
teleportation of continuous-variable systems.
We have calculated various correlation measures and studied
their dependence on the pulse shape and mid-frequency, the
frequency response of the devices, and the initial mean
photon number.
In particular, the results suggest that nonclassical
correlations rapidly decrease with increasing mean photon number,
which may drastically limit the effectively realizable non-classical
correlation in continuous-variable systems.
As a realization of a four-port device, we have considered
a dielectric plate of a Lorentz-type complex permittivity in more
detail.

Finally, it should be remembered that the fidelity and the
correlation indices are not strict entanglement measures.
Nevertheless, they may be helpful to find characteristic
dependences and to estimate limits of entanglement
transmission. Whereas for low-dimensional discrete-variable
systems such as qubits numerical methods can be used
to calculate the entanglement degradation exactly,
the exploding effort prevents one from applying them
to continuous-variable systems such as strongly
squeezed two-mode vacuum states. It has therefore been a great
challenge to find explicit entanglement measures,
at least for some classes of states such as Gaussian states.

%%%%%%%%%%%%%%%%%%%%%%%%%%%%%%%%%%%%%%%%%%%%%%%%%%%%%%%%%%%%%%%%%%%%%%%%%%%%
%%%%%%%%%%%%%%%%%%%%%%%%%%%%%%%%%%%%%%%%%%%%%%%%%%%%%%%%%%%%%%%%%%%%%%%%%%%%

\section*{Acknowledgments}
This work was supported by the Deutsche Forschungsgemeinschaft
and by the RFBR-BRFBR Grant No. 00-02-81023.
We thank S. Scheel for enlightening discussions.

%%%%%%%%%%%%%%%%%%%%%%%%%%%%%%%%%%%%%%%%%%%%%%%%%%%%%%%%%%%%%%%%%%%%%%%%%%%%%
{\appendix
\section{Input-output relations }
\label{ior}

The operator input-output relations at a dispersive and absorbing
four-port read \cite{Grun96,Patr99}
\begin{equation}
\label{A.00}
      \hat b_j(\omega )=\sum_{i=1}^{2}T_{ji}\left( \omega \right)
      \hat a_i\left(
      \omega \right)
      +\sum_{i=1}^{2}A_{ji}\left( \omega \right) \hat g_i\left( \omega \right),
      \end{equation}
where $\hat{a}_i(\omega)$ and
$\hat{b}_j(\omega)$  are respectively the
input- and output (destruction) operators of the radiation,
and $\hat{g}_i(\omega)$ are the
bosonic (destruction) operators of the device excitations.
The $2\times 2$ matrices ${\bf T}(\omega)$ (transformation matrix) and
${\bf A}(\omega)$ (absorption matrix), which are determined by
the complex refractive-index profile of the device \cite{Grun96},
satisfy the relation
\begin{equation}
\label{A.05}
      {\bf T}(\omega){\bf T}^+(\omega)
      + {\bf A}(\omega){\bf A}^+(\omega) ={\bf I} ,
      \end{equation}
provided that the device is embedded in vacuum.
Whereas the matrix ${\bf T}(\omega)$ describes the
effects of transmission and reflection, the matrix ${\bf A}(\omega)$
results from the material absorption.

Let
\begin{equation}
\label{A.09}
      \hat{\varrho}_{\rm in} =
      \hat{\varrho}_{\rm in}\big[\hat{\mbox{\boldmath $\alpha$}}(\omega),
      \hat{\mbox{\boldmath $\alpha$}}^\dagger(\omega)\big]
      \end{equation}
be the overall input density operator, which is an operator
functional of $\hat{\mbox{\boldmath $\alpha$}}(\omega)$ and
$\hat{\mbox{\boldmath $\alpha$}}^\dagger(\omega)$, with
$\hat{\mbox{\boldmath $\alpha$}}(\omega)$ being a
``four-vector'' according to
\begin{equation}
\label{A.08}
      \hat{\mbox{\boldmath $\alpha$}}(\omega)
      =\left[
      \hat{a}_1(\omega),\,
      \hat{a}_2(\omega),\,
      \hat{g}_1(\omega),\,
      \hat{g}_2(\omega)
      \right]^T.
      \end{equation}
The operator input-output relation (\ref{A.00}) is then
equivalent to the quantum-state transformation \cite{Knoe99}
\begin{equation}
\label{A.10}
      \hat{\varrho}_{\rm out}
      = \hat{\varrho}_{\rm in}
      \big[\mbox{\boldmath $\Lambda$}^+(\omega)
      \hat{\mbox{\boldmath $\alpha$}}(\omega),
      \mbox{\boldmath $\Lambda$}^T(\omega)
      \hat{\mbox{\boldmath $\alpha$}}^\dagger(\omega)\big].
      \end{equation}
The unitary $4\times 4$ matrix $\mbox{\boldmath $\Lambda$}(\omega)$
can be expressed in terms of the  $2\times 2$ matrices
${\bf T}(\omega)$ and ${\bf A}(\omega)$ as
\begin{equation}
\label{A.11}
      \mbox{\boldmath $\Lambda$}(\omega)
      =\left(\begin{array}{cc}{\bf T}(\omega)
      &{\bf A}(\omega)\\[1ex]
      {\bf F}(\omega)
      &{\bf G}(\omega)\end{array}\right),
      \end{equation}
where ${\bf F}(\omega)=-{\bf S}(\omega){\bf C}^{-1}(\omega){\bf
T}(\omega)$, ${\bf G}(\omega)={\bf C}(\omega){\bf S}^{-1}(\omega)
{\bf A}(\omega)$, and ${\bf C}(\omega) = \sqrt{{\bf T}(\omega)
{\bf T}^+(\omega)}$, ${\bf S}(\omega) = \sqrt{{\bf A}(\omega)
{\bf A}^+(\omega)}$ are commuting positive Hermitian matrices with
\begin{equation}
\label{A.14}
      {\bf C}^2(\omega) +  {\bf S}^2(\omega) = {\bf I} .
      \end{equation}

%%%%%%%%%%%%%%%%%%%%%%%%%%%%%%%%%%%%%%%%%%%%%%%%%%%%%%%%%%%%%%%%%%%%%%%%%%
%%%%%%%%%%%%%%%%%%%%%%%%%%%%%%%%%%%%%%%%%%%%%%%%%%%%%%%%%%%%%%%%%%%%%%%%%%

\section{Derivation of equation (\protect\ref{2.19})}
\label{app.chi2}

Combining equations (\ref{2.09}) and (\ref{2.17}) yields
\begin{eqnarray}
\label{C.01}
\lefteqn{
      \Phi_{\rm out} (\alpha ,\beta )
      = \left\langle \exp \left( \alpha
      \hat a^{\dagger }\!\left[\eta'\right] +
      \beta \hat d^{\dagger }\!\left[\tilde{\eta}'\right]-{\rm
      H.c.}\right)
      \right\rangle _{\rm out}
      }
      \nonumber\\&&\hspace{1ex}
      = \left\langle 0 \left|\hat{S}^{\prime \dagger }
      \left( q\right) \exp\!\left( \alpha
      \hat a^{\dagger }\!\left[\eta'\right] +
      \beta \hat d^{\dagger }\!\left[\tilde{\eta}'\right]-{\rm H.c.}
      \right)\hat S^{\prime }\left( q\right)
      \right| 0 \right\rangle
      \nonumber\\&&\hspace{1ex}
      = \left\langle 0 \left| \exp\!\left( \alpha
      \hat{S}^{\prime \dagger }\left( q\right)\hat a^{\dagger }
      \!\left[\eta'\right]
      \hat S^{\prime }\left( q\right) +
      \beta \hat{S}^{\prime \dagger }\left( q\right)\hat d^{\dagger }
      \!\left[\tilde{\eta}'\right]
      \hat S^{\prime }\left( q\right)-{\rm H.c.}\right)
      \right| 0 \right\rangle.
      \end{eqnarray}
It is not difficult to prove, on applying  the operator expansion
theorem, that
\begin{eqnarray}
\label{C.04}
\lefteqn{
      \hat{S}^{\prime \dagger }\left( q\right)
      \hat a\!\left[\eta'\right]
      \hat{S}^{\prime}\left( q\right) =
      }
      \nonumber\\&&\hspace{1ex}
      =\hat a\!\left[\eta'\right]
      -e^{i\varphi _q}\sinh |q|\|T\eta\|
      \left(\|\tilde{T}\tilde{\eta}\|\,\,
      \hat d^{\dagger }\!\left[\tilde{\eta}'\right]
      + (1-\|\tilde{T}\tilde{\eta}\|^2)^{1/2}\,
      \hat p^{\dagger }_{\tilde{\eta}} \right)
      \nonumber \\&&\hspace{1ex}
      + (\cosh |q|\!-\!1)\, \|T\eta\|\left(\|T\eta\|\,
      \hat a\!\left[\eta'\right]
      + (1-\|T\eta\|^2)^{1/2}\,\hat q_\eta \right)
      \end{eqnarray}
and
\begin{eqnarray}
\label{C.05}
\lefteqn{
      \hat{S}^{\prime \dagger }\left( q\right)
      \hat d\!\left[\tilde{\eta}'\right]
      \hat{S}^{\prime}\left( q\right) =
      }
      \nonumber\\&&\hspace{1ex}
      =\hat d\!\left[\tilde{\eta}'\right]
      -e^{i\varphi _q}\sinh |q|\|\tilde{T}\tilde{\eta}\|
      \left(\|T\eta\|\,\,
      \hat a^{\dagger }\!\left[\eta'\right]
      + (1-\|T\eta\|^2)^{1/2}\,\hat q^{\dagger }_\eta \right)
      \nonumber \\&&\hspace{1ex}
      + (\cosh |q|\! -\!1)\, \|\tilde{T}\tilde{\eta}\|
      \left(\|\tilde{T}\tilde{\eta}\|\,
      \hat d\!\left[\tilde{\eta}'\right]
      + (1-\|\tilde{T}\tilde{\eta}\|^2)^{1/2}\,
      \hat p_{\tilde{\eta}} \right).
      \end{eqnarray}
Substituting these expressions into equation (\ref{C.01})
and performing a normal ordering procedure, after some
straightforward calculations we arrive at $\Phi_{\rm out} (\alpha ,\beta )$
as given in equation (\ref{2.19}). Note that for
$T(\omega)=\tilde{T}(\omega) \equiv 1$ the characteristic function
$\Phi_{\rm in} (\alpha ,\beta )$ of the quantum state of the incoming
fields is obtained.

%%%%%%%%%%%%%%%%%%%%%%%%%%%%%%%%%%%%%%%%%%%%%%%%%%%%%%%%%%%%%%%%%%%%%%%%%%%%%
%%%%%%%%%%%%%%%%%%%%%%%%%%%%%%%%%%%%%%%%%%%%%%%%%%%%%%%%%%%%%%%%%%%%%%%%%%%%%

\section{Derivation of equation (\protect\ref{3.04})}
\label{app.F}

According to equations (\ref{2.08}), (\ref{2.09}), and  (\ref{3.03}),
we may write
\begin{eqnarray}
\label{B.01}
\lefteqn{
      F_e={\rm Tr}^{(a,d)}\left(\,\hat{\varrho}_{\rm in}^{(a,d)}
      \hat{\varrho}_{\rm out}^{(a,d)}\right)
      }
      \nonumber\\&&\hspace{1ex}
      = {\rm Tr}^{(a,d)}\left\{
      \hat{S}(q) |0_a,0_d\rangle
      \langle 0_a,0_d|\hat{S}^\dagger(q)
      {\rm Tr}^{(\cal{E})}\!\left[
      \hat{S}'(q) |0\rangle
      \langle 0|\hat{S}^{'\dagger}(q)
      \right]\right\}
      \nonumber\\&&\hspace{1ex}
      ={\rm Tr}\!\left\{\hat{S}^{'\dagger}(q)\hat{S}(q)
      |0_a,0_d\rangle\langle 0_a,0_d|\hat{S}^\dagger(q)
      \hat{S}'(q) |0\rangle\langle 0|\right\}
      \nonumber\\&&\hspace{1ex}
      =\langle 0|\hat{S}^{'\dagger}(q)\hat{S}(q)
      |0_a,0_d\rangle\langle 0_a,0_d|
      \hat{S}^\dagger(q)\hat{S}'(q)|0\rangle\,.
      \end{eqnarray}
We introduce the series expansions
\begin{eqnarray}
\label{B.02}
      \hat{S}(q) |0_a,0_d\rangle
      =\frac{1}{\cosh |q|} \sum_{n=0}^{\infty}
      {\tanh}^n |q| \,e^{-in\varphi _q}\,
      \frac{(\hat{a} ^{\dagger}[\eta] \hat{d}
      ^{\dagger}[\tilde{\eta}])^n}{n!}\,|0_a,0_d\rangle,
      \end{eqnarray}
\begin{eqnarray}
\label{B.03}
      \hat{S}' (q) |0\rangle
      =\frac{1}{\cosh |q|} \sum_{m=0}^{\infty}
      {\tanh}^m |q| \,e^{-im\varphi _q}\,
      \frac{(\hat{a}'^{\dagger}[\eta] \hat{d}'^{\dagger}
      [\tilde{\eta}])^m}{m!}\,|0\rangle
      \end{eqnarray}
and find
\begin{eqnarray}
\label{B.04}
\lefteqn{
      \langle 0_a,0_d|
      \hat{S}^\dagger\left(q\right)\hat{S}'\left(q\right)|0\rangle =
      \frac{1}{\cosh^2 |q|} \sum_{n=0}^{\infty}
      \bigg\{ {\tanh}^{2n} |q|
      }
      \nonumber\\&&\hspace{5ex} \times\,
      \langle 0_a,0_d|\,
      \frac{(\hat{a} [\eta] \hat{d}[\tilde{\eta}])^n}{n!}\,
      \frac{(\|T\eta\|\|\tilde{T}\tilde{\eta}\|\,
      \hat a ^{\dagger}\!\left[\eta'\right]
      \hat{d}^{\dagger}\!\left[\tilde{\eta}'\right])^n}{n!}\,
      |0\rangle\bigg\}.
      \end{eqnarray}
Combining equations (\ref{B.01}) and (\ref{B.04}) and using the
the commutation relations
\begin{eqnarray}
\label{B.05}
      \left[ \hat a[\eta] , \hat a^{\dagger }[\eta']
      \right] &=& (\eta | T \eta )/\|T\eta\|,
      \\
      \left[ \hat d[\tilde{\eta}] , \hat d^{\dagger }[\tilde{\eta}']
      \right] &=& (\tilde{\eta} | \tilde{T} \tilde{\eta} )/
      \|\tilde{T}\tilde{\eta}\|,
      \end{eqnarray}
we eventually arrive at
\begin{eqnarray}
\label{B.06}
      F_e &=& \frac 1{\cosh ^4|q|}\left| \sum_{m=0}^\infty
      \left[(\eta | T \eta )(\tilde{\eta}
      | \tilde{T} \tilde{\eta} )
      {\tanh }^2|q| \right]^m\right| ^2
      \nonumber \\&=&
      \frac {1}{\cosh ^4|q|}\cdot \frac {1}
      {\left| 1-{(\eta | T \eta )
      (\tilde{\eta} | \tilde{T} \tilde{\eta} )\tanh }^2|q|
      \right|^2}
      \nonumber \\&=&
      \left| 1+\left[1-(\eta | T \eta )(\tilde{\eta}
      | \tilde{T} \tilde{\eta} )\right]\,{\sinh }^2|q|
      \right| ^{-2}.
      \end{eqnarray}
}
%%%%%%%%%%%%%%%%%%%%%%%%%%%%%%%%%%%%%%%%%%%%%%%%%%%%%%%%%%%%%%%%%%%%%%%%%%%%%
%%%%%%%%%%%%%%%%%%%%%%%%%%%%%%%%%%%%%%%%%%%%%%%%%%%%%%%%%%%%%%%%%%%%%%%%%%%%%

\end{document}